# Single-Shot 3D Widefield Fluorescence Imaging with a Computational Miniature Mesoscope

Yujia Xue[1], Ian G. Davison[2,4], David A. Boas[1,3,4], Lei Tian[1,4, *]

[1]Department of Electrical and Computer Engineering, Boston University, Massachusetts 02215, USA.
[2]Department of Biology, Boston University, Massachusetts 02215, USA.
[3]Department of Biomedical Engineering, Boston University, Massachusetts 02215, USA.
[4]Neurophotonics Center, Boston University, Massachusetts 02215, USA.

[*]Corresponding author email: leitian@bu.edu

## Abstract

Fluorescence imaging is indispensable to biology and neuroscience. The need for large-scale imaging in freely behaving animals has further driven the development in miniaturized microscopes (miniscopes). However, conventional microscopes / miniscopes are inherently constrained by their limited space-bandwidth-product, shallow depth-of-field, and inability to resolve 3D distributed emitters. Here, we present a Computational Miniature Mesoscope ($CM^2$) that overcomes these bottlenecks and enables *single-shot* 3D imaging across an 8 × 7-$mm^2$ field-of-view and 2.5-mm depth-of-field, achieving 7-µm lateral resolution and better than 200-µm axial resolution. Notably, the $CM^2$ has a compact lightweight design that integrates a microlens array for imaging and an LED array for excitation in a single platform. Its expanded imaging capability is enabled by computational imaging that augments the optics by algorithms. We experimentally validate the mesoscopic 3D imaging capability on volumetrically distributed fluorescent beads and fibers. We further quantify the effects of bulk scattering and background fluorescence on phantom experiments.

**Introduction**

Fluorescence microscopy is an indispensable tool in fundamental biology and systems neuroscience (*1*). A major focus for recent technological developments is aimed at overcoming the barrier of *scale* (*2*). For example, perception and cognition arise from extended brain networks spanning millimeters to centimeters (*3*), yet rely on computations performed by individual neurons only a few micrometers in size (*4*). Recent progress, such as macroscopes (*3*), Mesolens microscope (*5*), two-photon mesoscope (*6*), RUSH (*7*), and COSMOS (*8*), are only beginning to bridge these scales. However, the development of such mesoscopic imaging systems is confounded by the scale-dependent geometric aberrations of optical elements (*9*). This results in an undesirable tradeoff between the achievable space-bandwidth-product (SBP) and the complexity of the optical design (*9, 10*), as evident by mesoscopes developed based on both the sequential (*5, 6*) and multiscale (*7*) lens design principles. In addition, the achievable field-of-view (FOV) is further constrained by the system's shallow depth-of-field (DOF) in many bioimaging applications (*3, 7*). For example, the FOVs for cortex-wide imaging systems are often set by the curved cortical surface that requires additional mechanisms to be compensated for, otherwise resulting in excessive out-of-focus blurs in the peripheral FOV regions (*3*).

Another technological focus is towards *miniaturization* driven by the need for long-term *in vivo* imaging in freely behaving animals. In particular, miniaturized head-mounted fluorescence microscopes, i.e. "miniscopes" (*11*), have made significant progress and enabled unprecedented access to neural signals, revealing new views of neural circuits underlying diverse behaviors, such as navigation, memory storage, learned motor programs, and social interactions. However, the imaging performance of current miniscope systems remains restricted by their optics, similar to their standard fluorescence microscopy counterparts. Most importantly, multiscale measurements are still beyond reach. Most of current miniscope systems limit imaging areas to under 1 mm$^2$ (*11*), confining measurements to a subset of cells within a single brain region. While larger FOVs are possible, fundamental physical limits preclude meeting the joint requirements of scale, resolution, and compactness by simply scaling up standard optical designs (*12*). In addition, widefield measurements only give access to fluorescence signals within a limited depth of several micrometers around the plane of focus, as set by the DOF of the optics (*11*). The head-mounted configuration further constrains the flexibility of adjusting focus, making imaging of 3D distributed emitters highly challenging (*13*). Although two-photon miniscopes have been developed to provide 3D scanning capability (*14, 15*), they require specialized optics and suffer from slow acquisition speed (*11*).

Here, we introduce and demonstrate a Computational Miniature Mesoscope (CM$^2$) that enables *large-scale 3D fluorescence* measurements with a *compact* and *lightweight* optical platform. The CM$^2$ uses simple optics and accomplishes its SBP improvement and 3D imaging capability without the need for mechanical scanning. It bypasses the physical limitations of the optics by jointly designing the hardware and the algorithm. Specifically, the CM$^2$ is capable of reconstructing 3D fluorescence distributions in 8.1 × 7.3 × 2.5 mm$^3$ volumes and achieving ~7-μm lateral resolution and better than 200-μm axial resolution from a *single* widefield measurement. This represents at least one order of magnitude increase in the FOV and two orders of magnitude improvement in the DOF over current miniscope systems, while still offering cellular level lateral resolution.

The imaging method of the CM$^2$ combines ideas from integral imaging (*16*), lightfield microscopy (*17–19*), compound-eye imaging (*20*), array microscopy (*21*, *22*), and coded aperture imaging (*23–26*). It works by first collecting a single 2D measurement using a microlens array (MLA), and then computationally reconstructing the 3D fluorescence distribution based on pre-characterized PSFs, as summarized in Fig. 1. Unlike systems designed to acquire 3D information by attaching an MLA to an existing microscope (*17–19*) or miniscope (*13*, *27*), the CM$^2$ uses the MLA as the *sole* imaging element (Figs. 1A and 1B), allowing our setup to circumvent the FOV limitations imposed by the conventional objective lens (*17*, *18*) or the GRIN lens (*13*, *27*). In addition, this configuration offers a simple and compact form factor by removing the bulk of infinite-conjugate optics used in existing miniscopes (*11*, *13*). Similar to coded-aperture techniques (*23–26*), the CM$^2$ captures 3D information through optical multiplexing, where the PSF is no longer a single tight focus but spreads over multiple foci. As compared to the techniques using highly dispersed PSFs (*23*, *24*), the CM$^2$ is designed to have a "small" 3 × 3 focal spot array that incurs a proportionally low degree of multiplexing, since both the image contrast and signal-to-noise ratio (SNR) degrade as multiplexing increases (*28*). This design ensures the CM$^2$ captures *high-contrast* measurements containing nine partially overlapping perspective projections from the object (see Figs. 1C). Furthermore, the *multi-view finite-conjugate* configuration provides the CM$^2$ with rapidly varying point spread functions (PSFs) across depths, which lays the foundation for *robust and accurate 3D reconstruction*. Accordingly, the forward model of the CM$^2$ describes the convolution between the 3D object and the depth-dependent PSFs simultaneously projected onto the image sensor. The CM$^2$ reconstruction algorithm recovers the 3D object by solving a sparsity-promoting regularized least-squares problem. As compared to the digital refocusing algorithm that synthesizes geometrically refocused images at different depths (*16*, *17*, *22*), the CM$^2$ algorithm provides *depth-resolved* reconstructions by solving the full 2D-to-3D deconvolution problem (Fig. 1D).

Importantly, the CM$^2$ operates as a standalone fluorescence imaging device that integrates the fluorescence excitation module with the imaging module on the same platform (Figs. 1A and 1B). Naively adopting the popular on-axis epi-illumination to a mesoscale FOV leads to bulky optics and undesired long working distance. Instead, we design and optimize an array of LEDs that create uniformly distributed illumination across a centimeter-scale FOV using an oblique epi-illumination configuration. In addition, this design imparts the compactness and light weight of the CM$^2$ and bypasses the conventional limitations from the collimating optics, dichromatic mirror (*11*), and diffusing elements needed for wide-FOV illumination (*12*) in existing miniscopes.

Building from off-the-shelf hardware components, a 3D printed housing, and augmented with the reconstruction algorithm, the CM$^2$ enables volumetric imaging of fluorescence objects and provides high resolution, wide FOV, and extended DOF with a 3D SBP of up-to 21.6 million voxels. Our joint optical-computational design allows us to perform tomographic reconstruction using a single measurement. In the following, we first outline the operation principle of the CM$^2$ and derive the theoretically achievable lateral and axial resolution based on both geometric optics and 3D modulation transfer function (MTF) analysis. We then show that our experimentally obtained resolution matches well with the theoretical predictions. Next, we experimentally demonstrate the 3D mesoscopy capability under different imaging conditions. First, we present results on scattering-free samples, including fluorescent particles embedded in clear volumes and fiber clusters spread over a curved surface. Next, we investigate the effects of bulk scattering and background fluorescence, and then quantify the axial reconstruction range of the CM² in a series of controlled phantom experiments. Finally, we demonstrate the advantage of the CM²'s extended DOF across a mesoscale FOV by imaging a scattering volume with a curved surface geometry.

**Results**

**Principle of the CM²**
The principle of the CM²'s single-shot 3D imaging capability can be explained by drawing an analogy to frequency division multiplexing (FDM). In the FDM, simultaneous transmission of multiple signals is made possible by first modulating each signal with a distinct carrier signal and later separating them by demodulation. Analogously, in the CM² simultaneously resolving fluorescence signals from multiple depths is achieved by first convolving the signals from each depth with a distinct PSF and later reconstructing the depth-wise information by deconvolution, as summarized in Fig. 2. Further considering fluorescence signals from a continuous volume, the axial resolving power of the CM² is fundamentally limited by the need for substantially differing PSFs across depths.

The experimentally measured PSFs by displacing a 5-µm source axially along the optical axis are shown in Fig. 2A (see details in Materials and Methods). The separation between the central and side foci decreases as the source is moved away from the MLA, resulting in the characteristic "axial shearing" in the array PSF. The separation of the foci $d$ as a function of the object-MLA distance $l$ (measured from the 1$^{st}$ principal plane $H_1$ of the MLA) can be quantified based on a geometric optics analysis, which gives

$$d = \left(\frac{l_0}{l} + 1\right)D \qquad (1)$$

where $D = 1$ mm is the pitch of the MLA and $l_0 = 5$ mm is the MLA-sensor distance (measured from the 2$^{nd}$ principal plane $H_2$ of the MLA). An excellent match between the model in Eq. (1) and the experiments are found by overlaying the theoretically predicted separation $d$ within the 2.5-mm defocus range onto the maximum intensity projection (MIP) of the experimentally measured PSFs in Fig. 2A.

The depth variation of the PSFs can be further quantified using the Pearson correlation coefficient (PCC) calculated between the nominal in-focus PSF (the $z = 0$ plane is defined at the nominal working distance $l = 12$ mm) and each defocus PSF (see Materials and Methods). As shown in Fig. 2A, the correlation reduces rapidly across different depth planes because of the axial shearing of the array focus. We further compare the 3 × 3 array PSF with other configurations, including a single focus, 2 × 3 array, and 2 × 2 array, and show that the decorrelation generally improves with the number of foci. To further quantitatively compare the axial resolving powers of different configurations, we adopt the full width at PCC = 0.8 proposed in (*26*) as a heuristic estimate of the axial resolution, which gives 155 µm for the 3 × 3 array, 179 µm for the 2 × 3 array, and 206 µm for the 2 × 2 array. In the next section, we show that these PCC-based axial resolution predictions agree well with our experimental data.

The achievable resolution of the CM² can be rigorously quantified by computing the system's 3D MTF. In Fig. 2B, the $f_x$-$f_z$ cross-sections of the 3D MTFs of the CM² and the single-microlens system are compared (see details in Materials and Methods). First, we analyze the axial resolution. Notably, the $CM^2$ dramatically *extends* the *axial* ($f_z$) *bandwidth*, and hence achieves much improved axial resolution. Akin to all microscopes (*29*), the axial bandwidth of the $CM^2$ depends on the lateral frequency ($f_x$). The high-frequency region, e.g. $f_x > 0.05$ µm$^{-1}$, supports a relatively uniform axial bandwidth of ~0.014 µm$^{-1}$, which predicts ~140-µm axial resolution. The low frequency region suffers from the common "missing-cone" problem (*29*), i.e. reduced axial Fourier coverage, which results in deteriorated axial sectioning. The shape of the MTF can be well explained by classical imaging theory. The diffraction-limited single microlens system has a 3D MTF with a "bowtie"-shaped $f_x$-$f_z$ cross-section (*29*), as illustrated in Fig. 2B. The axial bandwidth is $B_z^0 = \frac{NA^2}{\lambda} = 0.0033$ µm$^{-1}$, where the numerical aperture (NA) of the microlens is ~0.042 at the nominal working distance and the central emission wavelength $\lambda = 530$ nm. The extended axial Fourier coverage of the $CM^2$ is attributed to the array PSF. Specifically, each axially sheared side focus produces a *tilted* band through the origin in the 3D MTF, as illustrated in Fig. 2B. The tilting angle, $\alpha_{MLA} = \tan^{-1}(D/l) = 4.76°$, is set by the angular location of the side microlens, akin to the Fourier slice theorem (*30*), which agrees well with the experimental MTF. This MTF-based axial resolution analysis is further corroborated by a geometric optics analysis (see details in Materials and Methods). Together, these results show that the axial resolution of the $CM^2$ is fundamentally determined by the total angular coverage by the whole array. The MTF analysis further shows that the axial bandwidth of the current $CM^2$ prototype is ~1.56× narrower than the diffraction limit due to the pixel undersampling (see details in Materials and Methods), which in turn degrades the axial resolution proportionally.

Next, we analyze the lateral resolution. Due to the pixel undersampling, the MTF of the $CM^2$ only contains lateral frequency up to the sampling rate $f_s = \frac{1}{2\Delta} = 0.1$ µm$^{-1}$, where $\Delta = 5$ µm is the effective pixel size (see details in Materials and Methods). The MTF analysis shows that the $CM^2$ does not improve the lateral resolution as compared to the single-microlens case, since in both cases the lateral resolution is fundamentally limited by the NA of a *single* microlens. In addition, the lateral bandwidth of the $CM^2$ is not strongly affected by the axial frequency, indicating a relatively uniform lateral resolution regardless of the axial feature size. We later verify these predictions and demonstrate excellent agreement between the analysis and the experiments.

**Resolution and lateral shift variance characterization**
The image formation of the $CM^2$ is approximated by a slice-wise shift-invariant model. Under this approximation, the lateral shift *variance* at a given depth is neglected. Accordingly, the $CM^2$ measurement is modeled as the axial sum of the 2D convolution between each object slice at a given depth and the corresponding depth-dependent PSF (see details in Materials and Methods). This simplification is to reduce the requirements of both the physical PSF calibration and the computational complexity of the inversion algorithm. In practice, however, lateral shift variance is present due to several factors, including the spatially varying aberrations in the microlenses, the finite-sized image sensor, and the angle-dependent response of the CMOS pixel.

To characterize the lateral shift variance, we image a 5-µm pinhole across an 8 mm × 7 mm FOV. Next, we calculate the PCC between the on-axis PSF and each off-axis PSF to quantify the degree of lateral shift variance across the FOV in Fig. 2C (see details in Materials and Methods). Notably, both the finite sensor size and the limited pixel angular response can truncate the PSF to a smaller array from an off-axis point source, as shown in the examples in Fig. 2C. Accordingly, we divide the FOV into nine regions based on the number of foci in the PSF overlaid on the PCC map. The boundaries of these regions align well with where the PCC drops sharply, which shows that the loss of foci (i.e. "views") contributes significantly to the lateral shift variance. By further analyzing the PSF measurements, we conclude that the vertical PSF truncation is from the limited sensor size. The horizontal PSF truncation is from the limited angular pixel response, which has a $\sim 21°$ cutoff estimated from our measurements. Next, to assess the effect of the spatially varying aberrations of the microlens alone, we calculate the PCCs using only the focal spot from the central microlens. The PCC map in Fig. S4C shows that the central 7-mm diameter region displays high correlation (PCC > 0.7), indicating a relatively good match to our shift invariant model. Outside this region, the spatially varying aberrations further increases the system's lateral shift variance, which in turn degrades the reconstruction.

The impact of the lateral shift variance to the $CM^2$ resolution is quantified by deconvolving the pinhole measurements at different lateral locations using the same shift invariant model (see details in Materials and Methods). To account for the statistical variations, we aggregate the data into three groups in Fig. 2D, including those from the central (3 × 3 foci), outer (3 × 2 or 2 × 3 foci), and corner (2 × 2 foci) FOV regions (as defined in Fig. 2C), based on the number of foci/views in the measurement. The lateral resolution is consistently below 7 µm, as measured by the lateral full-width-at-half-maximum (FWHM) of the reconstructed intensity profile. Only slight variations are observed in the three FOV regions (central: 6.18 µm, outer: 6.34 µm, and corner: 6.49 µm). This result matches well with our MTF analysis which shows that the lateral resolution is not affected by the array size. The variations are likely due to the uncompensated off-axis aberrations at the peripheral FOV regions. On the other hand, the axial resolution (measured by the axial FWHM of the reconstructed intensity profile) is strongly affected by the number of views in the measurements, as the axial-PSF PCC analysis suggests. As validated in our experiment, the FWHM from the central 3 × 3-view region is ~139 µm, which matches well with the 3D MTF prediction; the outer 3 × 2-view and 2 × 3-view regions is ~172 µm; the corner 2 × 2-view region is ~189 µm. Overall, these results establish that the slice-wise shift-invariant model used in the current $CM^2$ prototype is sufficient to image an 8 mm × 7 mm FOV and provide 7-µm lateral resolution and better than 200-µm axial resolution.

**Experiment on a tilted fluorescent resolution target**
A fluorescent resolution target (Edmund Optics 57-895) is imaged to further validate the lateral resolution of the $CM^2$. We conduct experiments by tilting the target across the volume (Figs. 2E-G) or placing it on the same focal plane (Supplementary Materials Fig. S5A) and show that the same lateral resolution is achieved regardless of the geometry. The XY MIP of the $CM^2$ reconstruction (Fig. 2G) shows that the features with 7-µm linewidth (Group 6, Element 2) can be resolved, which agrees with both our MTF analysis and pinhole deconvolution results. The XZ MIP (Fig. 2G) demonstrates successful recovery of the tilted geometry. The feature size-dependent axial sectioning is also observed. The larger features result in wider axial elongations. Since we took the measurement by placing the high-resolution features in the central FOV region, the axial elongation artifacts are observed more prominently in the outer FOV regions corresponding to the low-resolution features on the target. We further validate these observations using Zemax simulated measurements (Supplementary Materials Fig. S5B and Section S6) and find good agreement between the simulations and the experiments.

**Experiments on fluorescent particles in clear volumes**
We experimentally demonstrate that the CM$^2$ allows localizing fluorescent emitters distributed across a large volume. First, we image 100-μm diameter fluorescent particles dispersed in a ~7.0 mm × 7.3 mm × 2.5 mm volume (the CM$^2$ measurement shown in Fig. 1C). We establish the accuracy of the CM$^2$ volumetric reconstruction result (Fig. 1D and Supplementary Materials Movie S1) by comparing it with an axial stack acquired using a 10× 0.25 NA objective lens on a commercial epi-fluorescence microscope (Nikon TE2000-U) in Supplementary Materials Fig. S8B and Section S10, where we show excellent agreement between the two. The axial elongation from each 100-μm particle is consistently around 420 μm.

Next, to test the CM$^2$'s performance on samples with the feature size similar to a single neuron, we image 10-μm diameter fluorescent particles distributed in a ~5.7 mm × 6.0 mm × 1.0 mm volume. The raw CM$^2$ measurement is contaminated with stronger background fluorescence due to the increased particle concentration and suffers from lower contrast and reduced SNR due to the reduced brightness of the emitters (shown in the insert in Fig. 3A). Nevertheless, the CM$^2$ algorithm is tolerant to these signal degradations, as shown in the high-quality full-FOV reconstruction in Fig. 3A. The mesoscale FOV offered by the CM$^2$ is highlighted by comparing that from the 2× and 10× objective lenses (on Nikon TE2000-U microscope with a PCO. Edge 5.5 sCMOS). Visual comparison between the CM$^2$ reconstruction and the axial stack acquired by the 10× 0.25 NA objective lens are shown in Fig. 3B, demonstrating accurate single-shot localization of the individual particles (volumetric visualization available in Supplementary Materials Movie S2). We further quantify the reconstruction accuracy by comparing the CM$^2$ reconstruction with the axial stacks taken with 2× 0.1 NA and 10× 0.25 NA objective lenses. As shown in the lateral cross-sections in Fig. 3C, the CM$^2$ accurately recovers the 10-μm particle profile. Further evaluating the axial cross-sections (Fig. 3C) indicates that the CM$^2$ reconstruction achieves better axial sectioning than the 0.1 NA objective lens but worse than the 0.25 NA objective lens. The XZ cross-sectional view of a single-particle reconstruction (Fig. 3D) also highlights these observations. Quantitatively, the axial elongation from a 10-μm particle taken from the central FOV is around 246 μm. To characterize the spatial variations of the reconstruction, we quantify the lateral sizes and axial elongations of all the reconstructed particles across the entire volume (see details in Materials and Methods). In Fig. 3E, the statistics of the lateral and axial FWHMs are computed for the central, outer, and corner FOV regions (as defined in Fig. 2C). For the lateral size, in both the central and outer FOV regions, the reconstructed particle widths are consistently around 11 μm. In the corner FOV region, the lateral size is broadened to an average 12.3 μm possibly due to the unmodeled stronger off-axis aberrations. On the other hand, the axial elongation is more affected by the missing-view induced lateral shift variance. In the central FOV region, the reconstructed particles have an average 246-μm elongation. In the outer and corner FOV regions, the axial widths are elongated to ~292 μm and ~299 μm, respectively. These results are in good agreement with those from our resolution characterization experiments.

**Experiment on fluorescent fibers on a curved surface**
The ability to image complex volumetric fluorescent samples is experimentally tested on fluorescent fibers spread on a 3D printed curved surface that mimics the surface profile of a mouse cortex (as shown in Fig. 4A). The sample spans a wide FOV (~7.8 mm × 4.9 mm) and an extended depth (~0.9 mm). As shown in the depth-color coded MIPs of the full-FOV reconstruction (Fig. 4B and Supplementary Materials Movie S3), the overall surface curvature with closely packed fiber structures can be clearly recovered. The reconstruction quality of the CM$^2$ reconstruction is highlighted by comparing a few reconstructed depths with the widefield fluorescence measurements using the 2× 0.1 NA and 10× 0.25 NA objective lenses (Fig. 4D). The CM$^2$ algorithm correctly recovers the in-focus structures and rejects the out-of-focus blurs in each depth, since it solves for the 3D object rather than mimicking the physical focusing on a microscope. We also plot the reconstruction cutline across a dense fiber cluster and compare it with the widefield measurement from the 2× 0.1 NA objective lens. The overlay verifies that the CM$^2$ resolves most of the individual fibers (Fig. 4C). The differences in the intensity of different fibers between the two cutlines are primarily due to the different illumination conditions used during the measurements (oblique epi-illumination for the CM$^2$ vs on-axis for the 2× objective lens on a standard epi-fluorescence microscope). Additional experiments on the same type of fluorescent fibers placed on a planar surface are conducted to further verify the above observations, as described in Supplementary Materials Section S9 and Fig. S8A.

**Experiments on controlled scattering phantoms**
To quantitatively evaluate the performance of the CM$^2$ under bulk scattering and strong background fluorescence, we conduct experiments on eight phantoms with progressively increasing scattering properties, as summarized in Fig. 5. All the phantoms are seeded with the same concentration of the target 25-μm fluorescent particles so that the differences in the reconstruction are only attributed to bulk scattering and background fluorescence. In addition, 1.1-μm background fluorescent particles are added to mimic unresolvable fluorescent sources commonly seen in biological samples (e.g. neuropils in the brain). The seeding density of the background fluorescent particles are kept the same for all the phantoms. This allows us to isolate the impact of volumetric scattering on the target fluorescent signals from the background fluorescent particles. To introduce bulk scattering, we seed non-fluorescent 1-μm polystyrene particles for mimicking refractive index inhomogeneities in tissues. Furthermore, the scattering strength is controlled with progressively higher seeding density. Specifically, the first phantom does not contain any additional scatterers, which serves as the benchmark. For the rest of the seven phantoms, we double the seeding density of the non-fluorescent scatterers in each sample (see more details in Materials and Methods). The scattering level for each phantom is quantified by the scattering mean free path $l_s$ (the yellow curve in Fig. 5B).

The raw CM$^2$ measurement is subject to strong background and reduced image contrast, as shown in the example image (the top left panel of Fig. 5A). To overcome this issue, we conduct a background subtraction procedure (as detailed in Materials and Methods) before performing the 3D deconvolution. This procedure can effectively remove the slowly varying background while maintaining high-fidelity signals from the high-contrast targets, as shown in the background removed image (the top right panel of Fig. 5A) and the overlay between the raw and background removed images (the bottom panel of Fig. 5A). To quantify the effects of background fluorescence and bulk scattering in the raw measurements, we calculate the signal-to-background ratio (SBR) (see details in Materials and Methods) and find that the average SBR is reduced from ~1.62 for the least scattering phantom ($l_s$ = 324 mm) to ~1.18 for the most scattering phantom ($l_s$ = 0.15 mm) in our experiments (the blue curve in Fig. 5B). Recall that an array PSF with more foci theoretically leads to reduced image contrast when strong background signals are present in the sample. To verify this, we quantify the spatial variations of the SBR for each phantom. Indeed, the SBRs averaged over the three sub-FOV regions (defined in Fig. 2C) show that the SBR consistently increases as the number of views in the measurement reduces for all the phantoms. Quantitatively, the local SBR increases on-average by ~0.053 from the central FOV (3 × 3 views) to the outer FOV (2 × 3 or 3 × 2 views) and by another ~0.10 in the corner FOV (2 × 2 views), as shown in Fig. 5B. Besides the scattering condition and the number of captured views, we find several other factors further influence the SBR, including the sample uniformity, the angle-dependent CMOS pixel response, and spatially varying aberrations of the microlenses.

We perform 3D reconstruction for each scattering phantom. All the deconvolutions are conducted using the same computational settings (i.e. the same set of regularization parameters and a fixed number of iterations), so that the influence from the (nonlinear) regularization terms can be considered approximately identical across all cases. The reconstruction results are visualized in the XZ MIPs in Fig. 5C. As the scattering increases, while it is still possible to resolve individual emitters, the reconstructed depth range gradually reduces. When the scattering is sufficiently strong, the $CM^2$ reconstructs the emitters within the superficial layer of the phantom (Phantoms 6-8). This observation is further quantified by measuring the reconstructed depth range in each case (the red curve in Fig. 5B) (more details in Materials and Methods). It is generally observed that the depth range reduces as the SBR reduces. The depth range is first limited by the background fluorescence (Phantoms 1-5). As the scattering increases, the range approaches the limit set by the single scattering mean free path (Phantoms 6-8), much like other widefield fluorescence techniques. When the scattering mean free path is shorter than the axial elongation from a 25-µm fluorescent particle (~372 µm), the experimentally measured depth range in Fig. 5B is set by this elongation due to the limited axial resolution (Phantoms 7-8). The estimated reconstruction depth range varies due to the surface variations present in each phantom, which is quantified by calculating the standard deviation in Fig. 5B. The variations are seen in the MIPs of the reconstructed volumes in Fig. 5C, where the white dashed line represents the estimated top surface of each phantom in the reconstruction.

**Experiment on a scattering sample with a curved surface geometry**
Although the reconstruction volume of the $CM^2$ is fundamentally limited by bulk scattering, next we show that it is still an effective solution of compensating for the surface curvature often present in a mesoscale FOV. To demonstrate this, we image a scattering ($l_s \sim 264$ µm) phantom with a curved surface geometry (Fig. 6A). The phantom is made using the same protocol as before. The entire surface spans approximately a 725-µm range. Although only the fluorescent emitters within the superficial layer can be recovered, the curvature of the surface is faithfully reconstructed by the $CM^2$, as highlighted in the reconstructed volume in Fig. 6B and visualized in Supplementary Materials Movie S4. The fidelity of the reconstruction is further validated against the widefield fluorescence measurements in Fig. 6C, which shows excellent agreement.

**Discussion**

In summary, a novel miniaturized fluorescence imaging system is demonstrated to enable single-shot mesoscopic 3D imaging. The $CM^2$ integrates the fluorescence imaging and the excitation modules on the same compact and lightweight platform. Simulations and experiments have been presented to establish the operation principle and 3D imaging capability of the $CM^2$. Its utility for 3D mesoscopic imaging under bulk scattering and strong background fluorescence has been experimentally quantitatively evaluated. This computational microscopy technique achieves a $cm^2$-scale FOV, a mm-scale DOF, ~7µm lateral resolution, and better than 200-µm axial resolution, and offers up-to 21.6 million-voxel information throughput in a single shot. Under bulk scattering, the CM$^2$ is still able to reliably reconstruct the fluorescence distribution in the superficial layer and digitally compensate for the curved surface geometry. With these unique combinations of imaging capabilities, we believe that this miniaturized system has a strong potential for achieving neural imaging on scales approaching the full extent of the mouse cortical surface with single-neuron level resolution, as an attractive alternative to the table-top one-photon macroscope systems (*3*, *7*, *8*). As a pilot study, we simulate a brain-wide imaging of vascular networks in Supplementary Materials Section S7 and Fig. S6. The results show promising results of imaging complex structures across a cortex-wide FOV and accommodating for mm-scale surface variations.

While the current CM$^2$ prototype is considerably more compact and lighter weight as compared to table-top systems for cortex-wide imaging (*3*, *7*, *8*), it is not yet compatible with head-mounted *in-vivo* applications. The size and weight of the current CM$^2$ prototype are primarily limited by the circuit boards of the image sensor (BFLY-PGE-50A2M-CS, FLIR) and the LED (LXML-PB01-0040, Lumileds), as detailed in Supplementary Materials Section S1. Miniature circuit boards will be developed in the future generations of CM$^2$ by incorporating advances made in the wearable miniscope systems (*11–13*). A preliminary estimate is made based on a miniature sensor board (MU9PC-MBRD, Ximea) for the same CMOS sensor chip that has been used in a head-mounted *in-vivo* system for mice (*32*) and a miniature LED board (LXZ1-PB01, Lumileds) that provides the same central wavelength and similar illumination flux. Based on this estimate, the total weight of the CM$^2$ can be reduced by more than 5× to under 4 grams. The total size can be reduced by more than 15× to ~13 mm × 13 mm × 10 mm (details in Supplementary Materials Section S1). With these additional efforts in miniaturization combined with the cortex-wide optical window implantation technologies (*3*), we envision that future generations of CM$^2$ can be an attractive head-mounted platform for full cortical *in-vivo* imaging in freely moving mice.

The imaging capability of the CM$^2$ can be further improved with future development in both hardware and algorithm. First, this first-generation device suffers from low light efficiency (~20% overall efficiency) due to the oblique epi-illumination geometry, limiting its application to weak fluorescent samples. A pilot study on imaging a Green fluorescent protein (GFP)-labeled mouse brain slice is described in Supplementary Materials Section S8 and Fig. S7, which demonstrates the mesoscopic imaging capability and the limitation of the current version of the CM$^2$. The future generations of CM$^2$ will improve the light efficiency by exploring alternative designs, such as using novel focusing optics (*12*), diffractive optical elements (*33*), or fiber optics-coupled light sources (*34*). Second, the imaging optics of the CM$^2$ is designed by heuristically balancing several hardware and imaging attributes, including the resolution, FOV, image contrast, and device complexity and size. Given this multi-dimensional design space and several intrinsic tradeoffs, it is highly possible that the imaging optics can be further optimized by using advanced computational procedures, such as those based on classical (e.g. the genetic algorithm (*35*)) or data-driven (e.g. machine learning (*36*, *37*)) algorithms. Here, we discuss several promising directions to pursue in the future. First, our analysis shows that both the lateral and axial resolution of this CM$^2$ prototype are primarily limited by the pixel undersampling. Practical solutions for alleviating the pixel undersampling include reducing the de-magnification factor (by decreasing the working distance) and using a CMOS sensor with a smaller pixel size. By further balancing the primary tradeoff for the FOV, the working distance and the sensor choice can be optimized for improved resolution. Second, the lateral resolution is fundamentally limited by the NA of a single microlens. Customized aperiodic microlens arrays (*19*, *26*) having non-identical NAs can open up a broader design space for improving the lateral resolution without compromising other imaging attributes. The axial resolution is physically limited by the total angular extent of the MLA, which can be moderately increased given the size constraint for the wearable application. Nevertheless, our computational imaging approach potentially allows leveraging advanced algorithms to overcome the limitations imposed by the physical optics. In particular, recent methods in deep learning (*38–41*) and spatio-temporal signal processing algorithms (*42*) can be adapted to the $CM^2$ to dramatically improve the axial resolution beyond the diffraction limit. Third, the FOV is currently limited by the computational model that neglects the lateral shift variance in the image formation. This limitation can be overcome by developing reconstruction algorithms that can effectively incorporate shift-variant PSFs, such as the local convolution model (*26*) and deep learning algorithms (*43*). As observed in our experiment, the FOV is ultimately limited by the limited angular pixel response (see Supplementary Materials Section S4 and Fig. S4B), which is 11.7 mm × 11.7 mm given the $\sim 20.9°$ cutoff from the standard CMOS sensor used in this $CM^2$ prototype. The emerging back-side illumination CMOS sensor is an appealing solution that can increase the angular range by more than 2× and will be investigated in future $CM^2$ platforms. Finally, the image contrast is primarily affected by the array size of the multi-focus PSF and the scattering conditions. It may be possible to improve the contrast by reducing the array size while maintaining the imaging resolution and FOV by optimizing the physical parameters of the MLA using advanced algorithms, such as the genetic algorithm (*35*) and deep learning (*36, 37*). In addition, advancing the illumination technology by incorporating the structured illumination (*44*) can be a promising solution to suppress the background fluorescence and improve the image contrast. With these improvements, we envision that future generations of $CM^2$ may open up new exciting opportunities in a wide range of large-scale *in-vivo* 3D neural recording and biomedical applications.

## Materials and Methods

### The CM² prototype

The CM² consists of two main parts for fluorescence imaging, including the imaging and illumination modules, as shown in Figs. 1A, and 1B. The detailed descriptions are provided in the Supplementary Materials Section S1, and Fig. S1. Briefly, for the imaging path, we choose an off-the-shelf MLA with rectangular apertures and 100% filling factor for the lens region (#630, Fresnel Technologies Inc., focal length = 3.3 mm, pitch = 1 mm, thickness = 3.3 mm). The MLA is first diced into a smaller array whose size is slightly larger than the 3 × 3 array (see Supplementary Materials Section S2 and Fig. S2). The extra size is needed to minimize vignetting due to the thickness of the MLA, as illustrated in the ray tracing in Zemax in Fig. S2. The 3 × 3 MLA is approximately centered about the image sensor (Aptina MT9P031, monochrome CMOS, sensor area 4.3 mm × 5.7 mm, 1944 × 2592 pixels, 2.2-μm pixel size, 8-bit image output, dynamic range 60 dB, dark noise $6.62e^-$). The back-surface of the MLA is held approximately 2.6 mm above from the image sensor by a 3D printed housing. The resulting finite-conjugate imaging system has a nominal working distance of ~12 mm away from the front-surface of the MLA. This design provides an overall ~2.4× de-magnification and a 5-μm effective pixel size at the object space, which is verified experimentally by imaging a resolution target. Compared with the 6.4-μm diffraction-limited lateral resolution, the CM² is undersampled by approximately a factor of 1.56 according to the Nyquist sampling requirement. No precise alignment is needed between the MLA and the image sensor. After assembly, a one-time system calibration is performed, in which a point source is scanned along the optical axis of the MLA to acquire a stack of PSFs (see Section PSF calibration and Supplementary Materials Section S3). A thin emission filter (535/50, Chroma Technology) is placed between the MLA and the sensor.

For the illumination path, four surface-mounted LEDs (LXML-PB01-0040, Lumileds) are placed symmetrically around the MLA for fluorescence excitation to provide oblique epi-illumination. The LEDs are connected to a driver (350 mA, 3021-D-E-350, LEDdynamics Inc.). Each LED is first filtered spectrally by the excitation filter (470/40, Chroma Technology), and then angularly confined by a 3D printed aperture to generate an oblique diverging beam for illuminating the imaging region. The positioning of the LEDs and the 3D printed apertures are optimized. To do so, we build a model in Zemax that incorporates the array geometry, the LED spectrum, the angular profile of each LED emitter, and the incident angle-dependent transmittance profiles of the excitation filter. We then optimize a merit function that considers the illumination uniformity and the total flux. Additional details about the illumination optimization are given in Supplementary Material Section S11. We achieve highly uniform illumination over a 10 mm × 10 mm area after the optimization, as shown in Fig. S9. The oblique geometry also reduces the transmittance efficiency of the emission filter, which is also modeled by our Zemax model (see Supplementary Material Section S12 and Fig. S10).

The CM² prototype is built around an off-the-shelf CMOS sensor package (BFLY-PGE-50A2M-CS, FLIR) by placing a 3D printed housing (material: black resin, printed on Form 2, FormLabs) on top of the sensor circuit board (Fig. 1B). Additionally, linear polarizing thin films (86-180, Edmund Optics) are inserted in front of each LED and the MLA. The orientation of each polarizer is adjusted to achieve optimal rejection of the specular reflections. After assembly, the prototype is connected to a desktop computer via an Ethernet cable for power and image acquisition. The prototype weighs ~19 grams and has a dimension of ~29 mm × 29 mm × 30 mm (see details in Supplementary Materials Section S1). Both the size and weight are primarily limited by the BFLY-PGE-50A2M-CS image sensor circuit board, which weighs 18 grams (~95% of the total weight) and has a dimension of ~29 mm × 29 mm × 23 mm. The lateral dimension is further limited by the LXML-PB01-0040 LED circuit board, which has a dimension of ~4.6 mm × 3.2 mm. Both a miniature image sensor board, MU9PC-MBRD (Ximea, the same Aptina MT9P031 CMOS chip, weight: 1.5 grams, dimension: ~13 mm × 13 mm × 3 mm) and miniature LED circuit board (LXZ1-PB01, Lumileds, dimension: 1.7 mm × 1.3 mm) will be incorporated in the future $CM^2$ platforms to further reduce the weight and size by following the protocol established in existing head-mounted platforms (*11*, *12*, *32*).

**PSF calibration**
After the $CM^2$ is assembled, it only requires a one-time calibration to characterize its PSFs. To perform the calibration, we first build a point source consisting of a green surface-mounted LED (M530L4, Thorlabs), diffused by multiple layers of highly scattering thin films (Parafilm), and followed by a 5-μm pinhole (P5D, Thorlabs). The details on the construction of the point source are in the Supplementary Material Section S3 and Fig. S3. The point source is mounted on a three-axis automatic translation stage and controlled by a custom-built MATLAB program. To calibrate the PSFs, the point source is scanned along the axial direction with a 10-μm step size across the [-3.5 mm - 3.5 mm] range. The measured PSFs are then registered numerically to account for the slight misalignment between the mechanical scanning axis and the optical axis. During the experiments, the central 2.5-mm range is used to perform the reconstruction. To quantify the lateral shift variance, we also measure the PSFs by scanning the pinhole across a 51 × 51 grid with a 0.2-mm step size over a 1-$cm^2$ FOV located at the working distance plane. During the experiments, the central 8 mm × 7 mm region is used to characterize the lateral shift variance of the system.

**3D MTF calculation**
After acquiring the system's axial PSF stack, we estimate the achievable resolution by computing the system's 3D MTF. Note that the MTF calculation assumes the system is spatially shift *invariant*. However, the $CM^2$ is shift *variant* at each focal plane due to the finite-sized image sensor and the angle-dependent response from the CMOS pixel that may truncate the array PSF, and the spatially varying aberrations of the microlenses. By neglecting these lateral shift variance effects, the 3D MTF is calculated by directly taking the 3D Fourier transform of the axial PSF stack.

**Axial resolution based on geometric optics analysis**
The axial resolution of the $CM^2$ is estimated based on the following geometric optics framework. By taking the finite difference on both sides of Eq. (1), it gives

$$\Delta d = \frac{Dl_0}{l^2} \Delta l \quad (2)$$

Equation (2) shows that an axial displacement by $\Delta l$ produces a lateral shift of the side focus by $\Delta d$. The smallest distinguishable lateral shift can be approximated by the pixel size, which in turn sets the *geometric-optics-limited* axial resolution. By setting $\Delta d = 5$ μm (the pixel size at the object space) and plugging in other physical parameters of our $CM^2$ prototype, Eq. (2) gives ~144μm axial resolution at the nominal working distance ($l = 12$ mm). This result matches well with the 3D MTF predicted 140-μm axial resolution.

**The CM² forward model and reconstruction algorithm**
The CM² is modeled by a slice-wise shift-*invariant* model. In this model, the 2D measurement is calculated as the axial sum of the 2D convolution between the object "slice" at each depth and the corresponding depth-dependent PSF. It further assumes an unknown boundary condition at the image plane by including a truncation operation to partially account for the loss of views at large incidence angles (*45*). At a given depth, the PSF also changes slowly across the FOV. However, to fully account for this lateral shift variance requires large costs from both the physical PSF calibration and computational reconstruction. We thus use the simplified slice-wise shift-invariant model that neglects the *lateral* shift variance. The degree of lateral shift variance is characterized in Fig. 2C-2D and Supplementary Materials Section S4 and Fig. S4. This simplification leads to slight degradation of the resolution (Fig. 2D) in the CM² reconstruction due to the model mismatch in the peripheral FOV regions (as shown in Fig. 2C-2D and Supplementary Materials Fig. S4). Concretely, the CM² forward model is written as the following compact form,

$$y = DHx \qquad (3)$$

where the discretized 3D object $x = [x_1, x_2, ..., x_n]^T$ is written as $n$ discrete depth slices that are concatenated into a long vector. The convolution operator $H = [H_1, H_2, ..., H_n]$ stacks all the corresponding 2D convolution matrices. The operation $Hx = \sum_{i=1}^{n} H_i x_i$ projects all the 2D measurements ($H_i x_i$) from different depths onto the 2D image sensor. $D$ is the truncation matrix.

The reconstruction algorithm solves an inverse problem that is highly ill-posed because of the dimensionality mismatch (i.e. from 2D to 3D). Our strategy is to incorporate priors by solving the following constrained optimization

$$\hat{x} = \arg\min_{x \geq 0} \frac{1}{2} \|DHx - y\|_2^2 + R(x) \qquad (4)$$

where $R$ includes two types of regularization terms, including the $l_1$-norm and the 3D total variation. The non-negativity constraint enforces the recovered fluorescent intensity to be positive and is achieved by minimizing an indicator function $1_+(\cdot)$. To efficiently solve this regularized least-squares problem, we adopt the alternating direction method of multipliers (ADMM) algorithm by splitting the problem into a sequence of sub-optimizations, where each sub-optimization has either a closed-form solution or a fast proximal operator (*45*). The iterative algorithm typically takes 0.5 and 2.5 hours to converge for 2D planar and 3D volumetric objects, respectively. The reconstruction algorithm is implemented in MATLAB 2018b and runs on the Boston University Shared Computing Cluster with an Intel Xeon Processor E5-2650 v2. The typical memory requirement is 16 GB and 256 GB for planar and volumetric object reconstructions, respectively. Additional details of the algorithm are provided in Supplementary Materials Section S5.

**Background subtraction algorithm**
We perform background subtraction on the raw CM² measurement to remove the slowly varying background before performing the 3D deconvolution, implemented in MATLAB 2018b. The background is estimated by applying the image morphological opening algorithm to the raw image, which is a two-step morphological operation. It first performs an image erosion followed by a dilation, both with the same template. In our case, the template is a disk with the diameter greater than the size of the fluorescent targets. It is observed that in the scattering phantom experiments, the signals from the emitters below the superficial layer are removed by this background subtraction procedure since these signals generally have low contrast and spread over a much wider area compared to those at the superficial layer.

**Calculation of the Pearson correlation coefficient (PCC)**
To quantitatively evaluate the depth variation of the array PSFs, we compute the PCC between the in-focus PSF and each defocus PSF in Fig. 2A. First, we register the defocus PSF with the in-focus PSF by computing the 2D cross-correlation. The location of the maximum in the cross-correlation map finds the lateral shift needed to register the pair of PSFs. Next, the PCC between the registered PSFs, $X$ and $Y$, is calculated by

$$\text{PCC}(X,Y) = \frac{(X-\mu_X)^T(Y-\mu_Y)}{\sqrt{\sum_i (X_i-\mu_X)^2}\sqrt{\sum_i (Y_i-\mu_Y)^2}} \quad (5)$$

where $\mu$ denotes the mean and $i$ is the pixel index of the registered PSFs. For other configurations (including the 2×3 array, 2×2 array, and single microlens), we first crop the original 3×3 array PSF to the desired array size (so the NA of the microlenses in each configuration is the same) and then repeat the above procedure to compute the PCCs.

To quantify the lateral shift variance of the $CM^2$, we compute the lateral PCC map (at the nominal in-focus plane $l = 12$mm) in Fig. 2C. We repeat the same cross-correlation-based registration between the on-axial PSF and each off-axis PSF and then compute the PCC. To further quantify the effect of spatially varying aberrations alone on the lateral shift variance, we compute the lateral PCC map of the focal spot from only the central microlens in Fig. S4C.

**Quantification of the resolution from pinhole measurements**
To quantify the resolution of the $CM^2$ at different lateral positions, we take measurements from a 5-μm pinhole that is scanned across an 8 mm × 7 mm FOV with 0.2-mm step size. The effect of the lateral shift variance on the reconstruction resolution is then evaluated by deconvolving the pinhole measurements using the shift invariant model. To reduce the computational cost, instead of deconvolving each measurement one by one, we perform deconvolution on a single synthetic image that is the sum of all the raw images. The volumetric reconstruction is performed by using the axial PSF stack (in Fig. 2A) and solving the slice-wise shift invariant deconvolution problem in Eq. (4). Following the reconstruction, we extract the 3D intensity profile for each point source. The lateral and axial resolution at each location is defined by the corresponding lateral and axial FWHMs of the intensity profile, respectively.

**Quantification of the reconstructed fluorescent particle size**
We use the following steps to quantify the lateral sizes and axial elongations of the deconvolved 10-μm fluorescent particles. First, each reconstructed particle is detected using the *3D object counter* tool in ImageJ. Next, the lateral and axial intensity profiles for each particle are extracted. Finally, the lateral size and axial elongation of each reconstructed particle are measured by the respective FWHMs of the intensity profile.

**Quantification of the signal-to-background ratio (SBR)**
For the experiments on phantoms with different scattering densities, we quantify the signal-to-background ratio (SBR) of the raw $CM^2$ measurements for each phantom. The SBR is calculated as the mean intensity value on the particle region over the mean intensity value on the background region. In our measurement, the background is not uniform across the whole FOV because different FOV regions capture different numbers of views (defined in Fig. 2C). To account for these variations, we divide the whole FOV into 500 μm × 500 μm FOV patches and quantify the local SBR of each patch in different FOV regions. The mean values of the local SBRs from the whole FOV under different scattering densities are plotted in the blue line in Fig. 5B along with the mean SBRs in different FOV regions labeled with different markers. When calculating the statistics, we remove the outliers from regions that contain bead clusters and boundary artifacts, such as glare. As expected, in the same FOV region, the SBR reduces as the scattering density increases. For a fixed scattering density, the SBR increases as the number of captured views reduces. This also matches with the visual inspection of the raw measurements shown in Fig. 5C. The SBR maps in Fig. 5B are formed by directly plotting the local SBR from each patch. The spatial variations in the SBR generally correlates well with the number of foci at the corresponding region (defined in Fig. 2C). Besides the scattering densities and number of foci, the SBR is further confounded by several experimental factors, including sample uniformity, the angular response of the CMOS pixels, and aberrations of the microlenses.

**Quantification of the reconstructed depth range**
To quantify the imaging depth limit of the $CM^2$ under bulk scattering, we measure the reconstructed depth range for each phantom. This is done by first randomly selecting multiple sub-FOV patches (800 μm × 800 μm) from the reconstructed volumes and then taking the XZ MIPs. We then use the *3D object counter* tool in ImageJ to detect the centroid of each reconstructed particle. For each sub-FOV, the local depth range is calculated as the difference between the maximum and minimum axial coordinates of the centroids with the additional axial elongation from each particle (372 μm for a 25-μm particle) to account for the intrinsic uncertainty in this measurement due to the limited axial resolution. Lastly, the mean and standard deviation are calculated from multiple sub-FOVs and reported in Fig. 5B. It is observed that the recovered depth range reduces as the SBR decreases. Additionally, the depth range curve crosses the scattering mean free path curve at Phantom 6 ($l_s$ = 497 μm) and later plateaus. This is because when the scattering mean free path (497 μm) is close to the axial resolution (372 μm), only emitters at the superficial layer of the sample can be faithfully reconstructed. When the scattering mean free path is less than the axial elongation from a single particle, the reconstructed depth range measured by our procedure is set by the axial blur induced by the imaging optics.

**Zemax simulation**
We conduct two series of simulations in Zemax, including the study of the imaging path and the illumination path of the $CM^2$. In the imaging path simulation, we use either a standard resolution target or a simulated mouse brain vasculature network as the sample. To make the simulation match the experimental conditions, the model incorporates the shift variant aberrations in the $CM^2$ by performing ray tracing in Zemax. All the components used in the setup, including the MLA, image sensor, and the 3D printed housing, are modeled to match the actual sizes in Zemax. The objects used in the simulation are first generated and then imported into the non-sequential mode of Zemax as source objects. The mouse brain vasculature object is generated by discretizing the volume into 16 discrete layers with a 0.1-mm layer thickness (details in Supplementary Materials Section S7 and Fig. S6). Furthermore, to account for the filter efficiency change under oblique illumination, we import the incidence dependent transmittance profiles of the filter set from the manufacturer (Fig. S10), as well as the emission and excitation spectra of the fluorophores used in the experiments. In the illumination path simulation, we optimize the uniformity and efficiency over a ~1 cm$^2$ excitation area. The surface-mounted LEDs, along with the 3D printed housing and filters, are accurately modeled in terms of their positions and spectral characteristics. A virtual detector is placed at the desired sample plane to measure the intensity profile of the excitation beam. Additional details are in Supplementary Materials Section S11 and Fig. S9.

**Calculation of the 3D space-bandwidth product (SBP)**
The SBP of a 3D imaging system (3D SBP) quantifies the fundamental spatial information throughput, which measures the maximum number of voxels that can be resolved inside the imaging volume in an ideal noise-free condition. In practice, to avoid possible ambiguities from different spatial resolution criteria, the SBP is calculated as the product between the imaging volume $V$ and the 3D bandwidth $B$ (*9*). Our experimentally measured maximum FOV and DOF are 8.1 mm × 7.3 mm and 2.5 mm, respectively, giving $V$ = FOV × DOF = 148 mm$^3$. The experimentally obtained lateral bandwidth $B_x$ is 0.143 µm$^{-1}$ (corresponding to 7-µm lateral resolution and a cutoff frequency $f_x$ = 0.0715 µm$^{-1}$), as marked in Fig. 2B. Further accounting for the square aperture of the microlens, the support of the $f_x$-$f_y$ cross-section is approximated as a square (i.e. the 2D autocorrelation of a square), whose area is $B_x^2/2$. The *local* axial bandwidth depends on the lateral frequency, as shown in the 3D MTF. Therefore, the practically achievable 3D SBP depends on the frequency content of the object. To give a reasonable estimate, we define the *upper bound* of the 3D SBP by using the axial bandwidth $B_z$ = 0.0143 µm$^{-1}$ at the lateral cutoff frequency $f_x$ = 0.0715 µm$^{-1}$, as marked in Fig. 2B. The 3D bandwidth is then estimated as $B = B_z B_x^2 / 2 = 1.46 \times 10^{-4}$ µm$^{-3}$, which approximates the irregular MTF shape (in Fig. 2B) as a rectangular volume. Accordingly, the *highest* 3D SBP of the CM$^2$ is approximately 21.6 million. The actual information throughput depends on the SNR of the measurements. The experimentally achievable SBP is further influenced by the sparsity of the object, the scattering condition, and the background fluorescence, as evident in the results presented in Fig. 5.

**Imaging of the fluorescent resolution target**
The resolution target needs to be excited at 365 nm and emits at 550 nm, which does not match our choice of the LEDs in the CM$^2$ prototype (designed for exciting common Green Fluorescence Proteins). As a result, the measurements of the target are taken with an external UV lamp.

**Scattering-free sample preparation**
Fluorescent particles with different sizes (10 µm and 100 µm, Thermo Scientific Fluoro-Max Green Dry Fluorescent Particles) are first suspended in clear resin (FormLabs, #RS-F2-GPCL-04) and then diluted to different concentrations. Next, we apply the mixture onto a standard 1-inch microscope slide. The samples are later cured under a UV lamp. The samples are controlled to be within 1~2 mm in thickness. The fluorescent fiber sample is made by soaking lens tissue fibers in green fluorescent dyes and then cured inside the clear resin. To mimic the surface curvature of the mouse cerebral cortex, fluorescent fibers are placed on top of a 3D printed clear mouse brain model that forms a total depth range of around 1 mm.

**Scattering phantom preparation**

We fabricate scattering phantoms with both bulk scattering and background fluorescence. The bulk scattering is controlled by embedding 1-µm non-fluorescent polystyrene microspheres (i.e. scatterers) (Thermo Scientific, 5000 Series Polymer Particle Suspension, refractive index = 1.5979) into the phantom. The background fluorescence is introduced by 1.1-µm green fluorescent microspheres (Thermo Scientific, Fluoro-Max Dyed Green Aqueous Fluorescent Particles) at a fixed density of $1.2 \times 10^6$ particles / mL. The imaging targets are 25-µm green fluorescent microspheres (Thermo Scientific, Fluoro-Max Green Dry Fluorescent Particles) of a fixed density of $1.5 \times 10^4$ particles / mL. The background medium is the clear resin (FormLabs, #RS-F2-GPCL-04, refractive index is approximately 1.5403) for the ease of fabrication. One caveat of this recipe is that the anisotropy factor $g$ of the phantoms is 0.965 due to the small refractive index contrast, which is larger than the commonly reported values for biological tissues (~0.9). This can result in worse background fluorescence in the raw measurements as compared to the case with smaller $g$ values (*31*).

Different amounts of non-fluorescent scatterers are added to the eight different phantoms with a micropipette (Thermo Scientific, Fisherbrand Elite Adjustable Volume Pipette, #FBE00100). Specifically, 0, 8, 16, 32, 64, 128, 256, and 512 µL of scatterer suspension (10% volume concentration) are added to 2 mL of clear resin, where 0 stands for the control "non-fluorescent scatterer-free" phantom. Correspondingly, the rest of the seven phantoms contain $7.6 \times 10^8$, $1.5 \times 10^9$, $3.0 \times 10^9$, $5.9 \times 10^9$, $1.2 \times 10^{10}$, $2.2 \times 10^{10}$, and $3.9 \times 10^{10}$ particles / mL, respectively. After fully mixing the bead suspension with the clear resin, 0.1 mL from each mixed solution is then transferred to a 3D printed well (inner diameter 8 mm, height 2 mm, clear resin). Each phantom is then cured under a UV lamp. The pictures of the phantoms used in our experiments are shown in Fig. 5D.

The scattering mean free path $l_s$ of each phantom is estimated based on Eq. (6) derived from the Mie scattering theory (*46*):

$$l_s = \frac{2d}{3\Phi Q_s} \qquad (6)$$

where $d$ is the mean diameter of the scatterers, and $\Phi$ is the volume fraction of the scatterers (calculated from the number of scatterers added to each phantom). $Q_s$ is the scattering efficiency factor calculated based on the Mie scattering calculator (*47*). For Phantom 1, we consider the 1.1-µm fluorescent beads as the main source of scattering and the corresponding $Q_s$ is 0.271. For Phantoms 2-8, we consider the 1.0-µm non-fluorescent beads as the main source of scattering and the corresponding $Q_s$ is 0.224. Accordingly, the scattering mean free paths $l_s$ for the eight phantoms are approximately 323.7 mm, 7.51 mm, 3.77 mm, 1.9 mm, 0.965 mm, 0.497 mm, 0.264 mm, and 0.147 mm, respectively.

# References

1. J. W. Lichtman, J.-A. Conchello, Fluorescence microscopy. *Nat. Methods*. **2**, 910–919 (2005).

2. S. Weisenburger, A. Vaziri, A Guide to Emerging Technologies for Large-Scale and Whole-Brain Optical Imaging of Neuronal Activity. *Annu. Rev. Neurosci.* **41**, 431–452 (2018).

3. T. H. Kim, Y. Zhang, J. Lecoq, J. C. Jung, J. Li, H. Zeng, C. M. Niell, M. J. Schnitzer, Long-Term Optical Access to an Estimated One Million Neurons in the Live Mouse Cortex. *Cell Rep.* **17**, 3385–3394 (2016).

4. C. Koch, Computation and the single neuron. *Nature*. **385**, 207–210 (1997).

5. G. McConnell, J. Trägårdh, R. Amor, J. Dempster, E. Reid, W. B. Amos, A novel optical microscope for imaging large embryos and tissue volumes with sub-cellular resolution throughout. *eLife*. **5**, e18659 (2016).

6. N. J. Sofroniew, D. Flickinger, J. King, K. Svoboda, A large field of view two-photon mesoscope with subcellular resolution for in vivo imaging. *eLife*. **5**, e14472 (2016).

7. J. Fan, J. Suo, J. Wu, H. Xie, Y. Shen, F. Chen, G. Wang, L. Cao, G. Jin, Q. He, T. Li, G. Luan, L. Kong, Z. Zheng, Q. Dai, Video-rate imaging of biological dynamics at centimetre scale and micrometre resolution. *Nat. Photonics*. **13**, 809–816 (2019).

8. I. V. Kauvar, T. A. Machado, E. Yuen, J. Kochalka, M. Choi, W. E. Allen, G. Wetzstein, K. Deisseroth, Cortical Observation by Synchronous Multifocal Optical Sampling Reveals Widespread Population Encoding of Actions. *Neuron*, S0896627320303159 (2020).

9. A. W. Lohmann, R. G. Dorsch, D. Mendlovic, Z. Zalevsky, C. Ferreira, Space–bandwidth product of optical signals and systems. *JOSA A*. **13**, 470–473 (1996).

10. D. J. Brady, N. Hagen, Multiscale lens design. *Opt. Express*. **17**, 10659–10674 (2009).

11. D. Aharoni, B. S. Khakh, A. J. Silva, P. Golshani, All the light that we can see: a new era in miniaturized microscopy. *Nat. Methods*. **16**, 11–13 (2019).

12. B. B. Scott, S. Y. Thiberge, C. Guo, D. G. R. Tervo, C. D. Brody, A. Y. Karpova, D. W. Tank, Imaging Cortical Dynamics in GCaMP Transgenic Rats with a Head-Mounted Widefield Macroscope. *Neuron*. **100**, 1045-1058.e5 (2018).

13. O. Skocek, T. Nöbauer, L. Weilguny, F. M. Traub, C. N. Xia, M. I. Molodtsov, A. Grama, M. Yamagata, D. Aharoni, D. D. Cox, P. Golshani, A. Vaziri, High-speed volumetric imaging of neuronal activity in freely moving rodents. *Nat. Methods*. **15**, 429–432 (2018).

14. W. Zong, R. Wu, M. Li, Y. Hu, Y. Li, J. Li, H. Rong, H. Wu, Y. Xu, Y. Lu, H. Jia, M. Fan, Z. Zhou, Y. Zhang, A. Wang, L. Chen, H. Cheng, Fast high-resolution miniature two-photon microscopy for brain imaging in freely behaving mice. *Nat. Methods*. **14**, 713–719 (2017).

15. B. N. Ozbay, G. L. Futia, M. Ma, V. M. Bright, J. T. Gopinath, E. G. Hughes, D. Restrepo, E. A. Gibson, Three dimensional two-photon brain imaging in freely moving mice using a miniature fiber coupled microscope with active axial-scanning. *Sci. Rep.* **8**, 1–14 (2018).

16. A. Stern, B. Javidi, Three-Dimensional Image Sensing, Visualization, and Processing Using Integral Imaging. *Proc. IEEE*. **94**, 591–607 (2006).

17. M. Levoy, R. Ng, A. Adams, M. Footer, M. Horowitz, Light Field Microscopy. *ACM SIGGRAPH 2006 Pap.*, 924–934 (2006).

18. A. Llavador, J. Garcia-Sucerquia, E. Sánchez-Ortiga, G. Saavedra, M. Martinez-Corral, View images with unprecedented resolution in integral microscopy. *OSA Contin.* **1**, 40–47 (2018).

19. L. Cong, Z. Wang, Y. Chai, W. Hang, C. Shang, W. Yang, L. Bai, J. Du, K. Wang, Q. Wen, Rapid whole brain imaging of neural activity in freely behaving larval zebrafish (Danio rerio). *eLife*. **6**, e28158 (2017).

20. J. Tanida, T. Kumagai, K. Yamada, S. Miyatake, K. Ishida, T. Morimoto, N. Kondou, D. Miyazaki, Y. Ichioka, Thin observation module by bound optics (TOMBO): concept and experimental verification. *Appl. Opt.* **40**, 1806–1813 (2001).

21. B. McCall, R. J. Olsen, N. J. Nelles, D. L. Williams, K. Jackson, R. Richards-Kortum, E. A. Graviss, T. S. Tkaczyk, Evaluation of a miniature microscope objective designed for fluorescence array microscopy detection of Mycobacterium tuberculosis. *Arch. Pathol. Lab. Med.* **138**, 379–389 (2014).

22. A. Orth, K. Crozier, Microscopy with microlens arrays: high throughput, high resolution and light-field imaging. *Opt. Express*. **20**, 13522–13531 (2012).

23. J. K. Adams, V. Boominathan, B. W. Avants, D. G. Vercosa, F. Ye, R. G. Baraniuk, J. T. Robinson, A. Veeraraghavan, Single-frame 3D fluorescence microscopy with ultraminiature lensless FlatScope. *Sci. Adv.* **3**, e1701548 (2017).

24. N. Antipa, G. Kuo, R. Heckel, B. Mildenhall, E. Bostan, R. Ng, L. Waller, DiffuserCam: lensless single-exposure 3D imaging. *Optica*. **5**, 1–9 (2018).

25. J. Shin, D. N. Tran, J. R. Stroud, S. Chin, T. D. Tran, M. A. Foster, A minimally invasive lens-free computational microendoscope. *Sci. Adv.* **5**, eaaw5595 (2019).

26. G. Kuo, F. Linda Liu, I. Grossrubatscher, R. Ng, L. Waller, On-chip fluorescence microscopy with a random microlens diffuser. *Opt. Express*. **28**, 8384 (2020).

27. Y. Chen, B. Xiong, Y. Xue, X. Jin, J. Greene, L. Tian, Design of a high-resolution light field miniscope for volumetric imaging in scattering tissue. *Biomed. Opt. Express*. **11**, 1662–1678 (2020).

28. S. R. Gottesman, E. E. Fenimore, New family of binary arrays for coded aperture imaging. *Appl. Opt.* **28**, 4344–4352 (1989).

29. C. J. R. Sheppard, X. Q. Mao, Three-dimensional imaging in a microscope. *J. Opt. Soc. Am. A*. **6**, 1260 (1989).

30. R. Ng, in *ACM SIGGRAPH 2005 Papers* (ACM, New York, NY, USA, 2005; http://doi.acm.org/10.1145/1186822.1073256), *SIGGRAPH '05*, pp. 735–744.

31. X. Cheng, Y. Li, J. Mertz, S. Sakadžić, A. Devor, D. A. Boas, L. Tian, Development of a beam propagation method to simulate the point spread function degradation in scattering media. *Opt. Lett.* **44**, 4989 (2019).

32. A. D. Jacob, A. I. Ramsaran, A. J. Mocle, L. M. Tran, C. Yan, P. W. Frankland, S. A. Josselyn, A compact head-mounted endoscope for in vivo calcium imaging in Freely behaving mice. *Curr. Protoc. Neurosci.* **84**, e51 (2018).

33. Z. Chen, B. M. Larney, J. Rebling, X. L. Deán-Ben, Q. Zhou, S. Gottschalk, D. Razansky, High-speed large-field multifocal illumination fluorescence microscopy. *Laser Photonics Rev.* **14**, 1900070 (2020).

34. A. Ray, M. A. Khalid, A. Demčenko, M. Daloglu, D. Tseng, J. Reboud, J. M. Cooper, A. Ozcan, Holographic detection of nanoparticles using acoustically actuated nanolenses. *Nat. Commun.* **11**, 171 (2020).

35. D. B. Conkey, A. N. Brown, A. M. Caravaca-Aguirre, R. Piestun, Genetic algorithm optimization for focusing through turbid media in noisy environments. *Opt. Express*. **20**, 4840–4849 (2012).

36. A. Durand, T. Wiesner, M.-A. Gardner, L.-É. Robitaille, A. Bilodeau, C. Gagné, P. De Koninck, F. Lavoie-Cardinal, A machine learning approach for online automated optimization of super-resolution optical microscopy. *Nat. Commun.* **9**, 5247 (2018).

37. Y. Wu, V. Boominathan, H. Chen, A. Sankaranarayanan, A. Veeraraghavan, in *2019 IEEE International Conference on Computational Photography (ICCP)* (2019), pp. 1–12.

38. M. Weigert, U. Schmidt, T. Boothe, A. Müller, A. Dibrov, A. Jain, B. Wilhelm, D. Schmidt, C. Broaddus, S. Culley, M. Rocha-Martins, F. Segovia-Miranda, C. Norden, R. Henriques, M. Zerial, M. Solimena, J. Rink, P. Tomancak, L. Royer, F. Jug, E. W. Myers, Content-aware image restoration: pushing the limits of fluorescence microscopy. *Nat. Methods*. **15**, 1090–1097 (2018).

39. Y. Xue, S. Cheng, Y. Li, L. Tian, Reliable deep-learning-based phase imaging with uncertainty quantification. *Optica*. **6**, 618–629 (2019).

40. Y. Wu, Y. Rivenson, H. Wang, Y. Luo, E. Ben-David, L. A. Bentolila, C. Pritz, A. Ozcan, Three-dimensional virtual refocusing of fluorescence microscopy images using deep learning. *Nat. Methods*, 1–9 (2019).

41. Y. Li, Y. Xue, L. Tian, Deep speckle correlation: a deep learning approach toward scalable imaging through scattering media. *Optica*. **5**, 1181–1190 (2018).

42. T. Nöbauer, O. Skocek, A. J. Pernía-Andrade, L. Weilguny, F. M. Traub, M. I. Molodtsov, A. Vaziri, Video rate volumetric $Ca^{2+}$ imaging across cortex using seeded iterative demixing (SID) microscopy. *Nat. Methods*. **14**, 811–818 (2017).

43. F. Xu, D. Ma, K. P. MacPherson, S. Liu, Y. Bu, Y. Wang, Y. Tang, C. Bi, T. Kwok, A. A. Chubykin, P. Yin, S. Calve, G. E. Landreth, F. Huang, Three-dimensional nanoscopy of whole cells and tissues with in situ point spread function retrieval. *Nat. Methods*. **17**, 531–540 (2020).

44. J. Mertz, Optical sectioning microscopy with planar or structured illumination. *Nat. Methods*. **8**, 811–819 (2011).

45. M. S. C. Almeida, M. Figueiredo, Deconvolving images with unknown boundaries using the alternating direction method of multipliers. *IEEE Trans. Image Process.* **22**, 3074–3086 (2013).

46. O. Mengual, G. Meunier, I. Cayré, K. Puech, P. Snabre, TURBISCAN MA 2000: multiple light scattering measurement for concentrated emulsion and suspension instability analysis. *Talanta*. **50**, 445–456 (1999).

47. Mie Scattering Calculator, (available at https://omlc.org/calc/mie_calc.html).

**Acknowledgments**

We thank Dr. Alberto Cruz-Martín for proving the brain slice, Dr. Kıvılcım Kılıç for helpful discussion on cortex-wide imaging, Daniel Leman for helping with the fluorescence filters and LED sources, Paul Mak for assisting with the microlens array dicing, Dr. Xiaojun Cheng and Waleed Tahir for helpful discussion on scattering experiments, Boston University Photonics Center for providing the access to Zemax, Boston University Shared Computing Cluster for proving the computational resources.

**Funding:** This work was supported by the National Eye Institute (NEI) (R21EY030016) and Boston University Dean's Catalyst Award.

**Author contributions:** L.T. and D.B. conceived the miniature mesoscope idea. Y.X. and L.T. designed the initial mesoscope system. Y. X., L. T. and I.D. further discussed and refined the design. Y.X. carried out the simulations and experiments. All authors contributed to the writing of the manuscript.

**Competing interests:** The authors declare that they have no competing interests.

**Data and materials availability:** All data needed to evaluate the conclusions in the paper are present in the paper and/or the Supplementary Materials. The data that support the findings in this study are available upon request from the corresponding author.

**Figures and Tables**

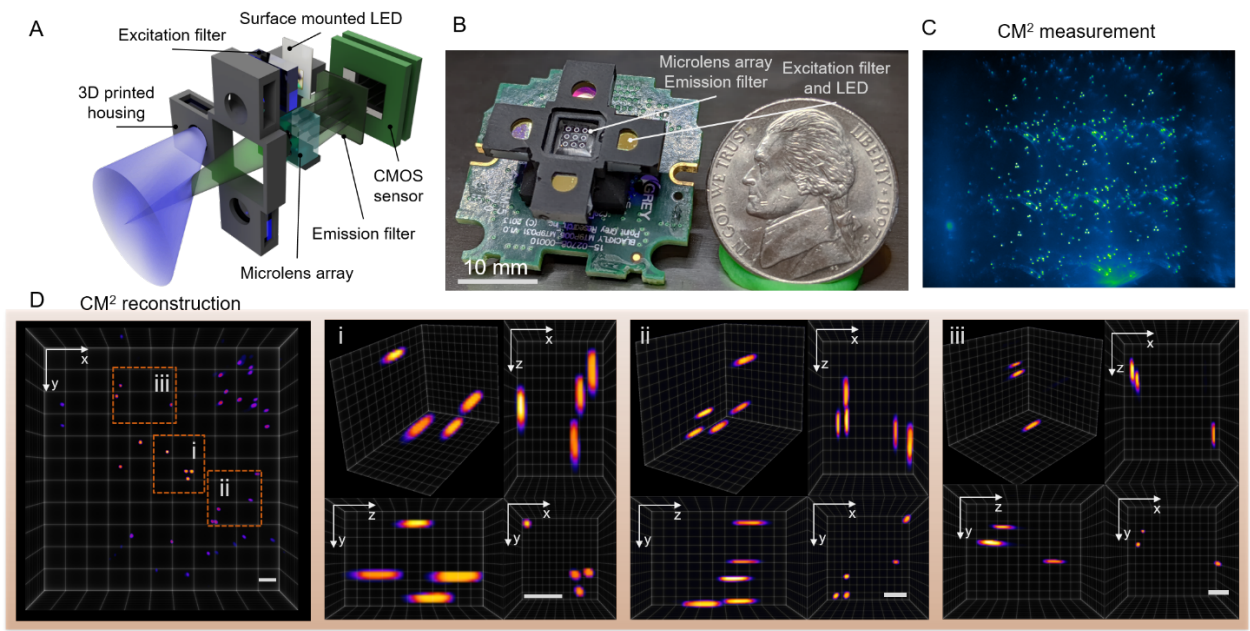

**Fig. 1. Single-shot 3D fluorescence Computational Miniature Mesoscope (CM$^2$).**
(**A**) The CM$^2$ combines a microlens array (MLA) optics and LED array excitation in a compact and lightweight platform. (**B**) A picture of the CM$^2$ prototype (the electric wires and the sensor driver are omitted). Photo Credit: Yujia Xue, Boston University. (**C**) The CM$^2$ measurement on 100-μm fluorescent particles suspended in clear resin. (**D**) The projected view of the CM$^2$ reconstructed volume (7.0 mm × 7.3 mm × 2.5 mm) and three zoom-in regions with orthogonal views. Scale bars: 500 μm.

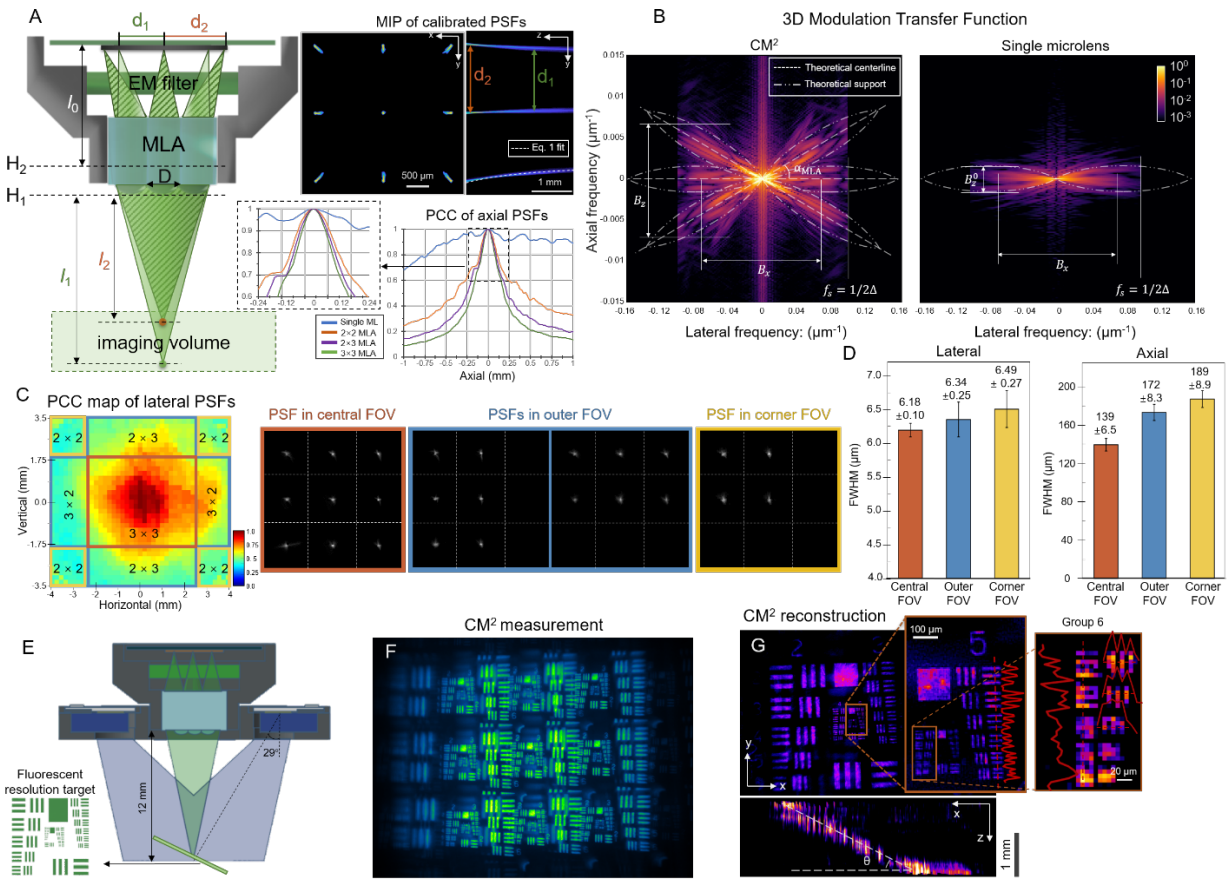

**Fig. 2. Characterization of the CM$^2$'s imaging principle, shift-variance, and resolution.** (**A**) The CM$^2$ produces axially varying array PSFs to achieve optical sectioning. The axial shearing in the side foci is well characterized by the geometric model in Eq. (1). The PCC of the axially scanned PSFs quantifies the expected axial resolution. (**B**) The 3D MTF (shown in log-scale) shows that the CM$^2$ captures extended axial frequency information and enlarges the system's SBP. The support of the experimental MTF matches with the theory (in dashed curve). The angle of each tilted "band" in the MTF is set by the angular location of the corresponding microlens $\alpha_{\text{MLA}}$ (in dash-dotted line). (**C**) The lateral shift variance is characterized by the PCC of the laterally scanned PSFs. The PSF in the central FOV (marked by orange boundary lines) contains 3 × 3 foci; the PSF in the outer FOV (marked by blue boundary lines) contains 2 × 3 or 3 × 2 foci; the PSF in the corner FOV (marked by yellow boundary lines) contains 2 × 2 foci. (**D**) The resolution at different regions of the FOV is characterized by reconstructing a 5-μm pinhole object using the CM$^2$'s shift invariant model. The lateral FWHM is consistently below 7 μm. The axial FWHM is ~139 μm in the central FOV and degrades to ~172 μm and ~189 μm in the outer and corner FOVs, respectively. (**E**) The geometry for imaging a tilted fluorescent target. (**F**) The raw CM$^2$ measurement. (**G**) The MIPs of the reconstructed volume (8.1 mm × 5.5 mm × 1.8 mm). The 7-μm features (Group 6, Element 2) can be resolved as shown in the zoom-in XY projection. The axial sectioning capability is characterized by the XZ projection, validating the feature size-dependent axial resolution.

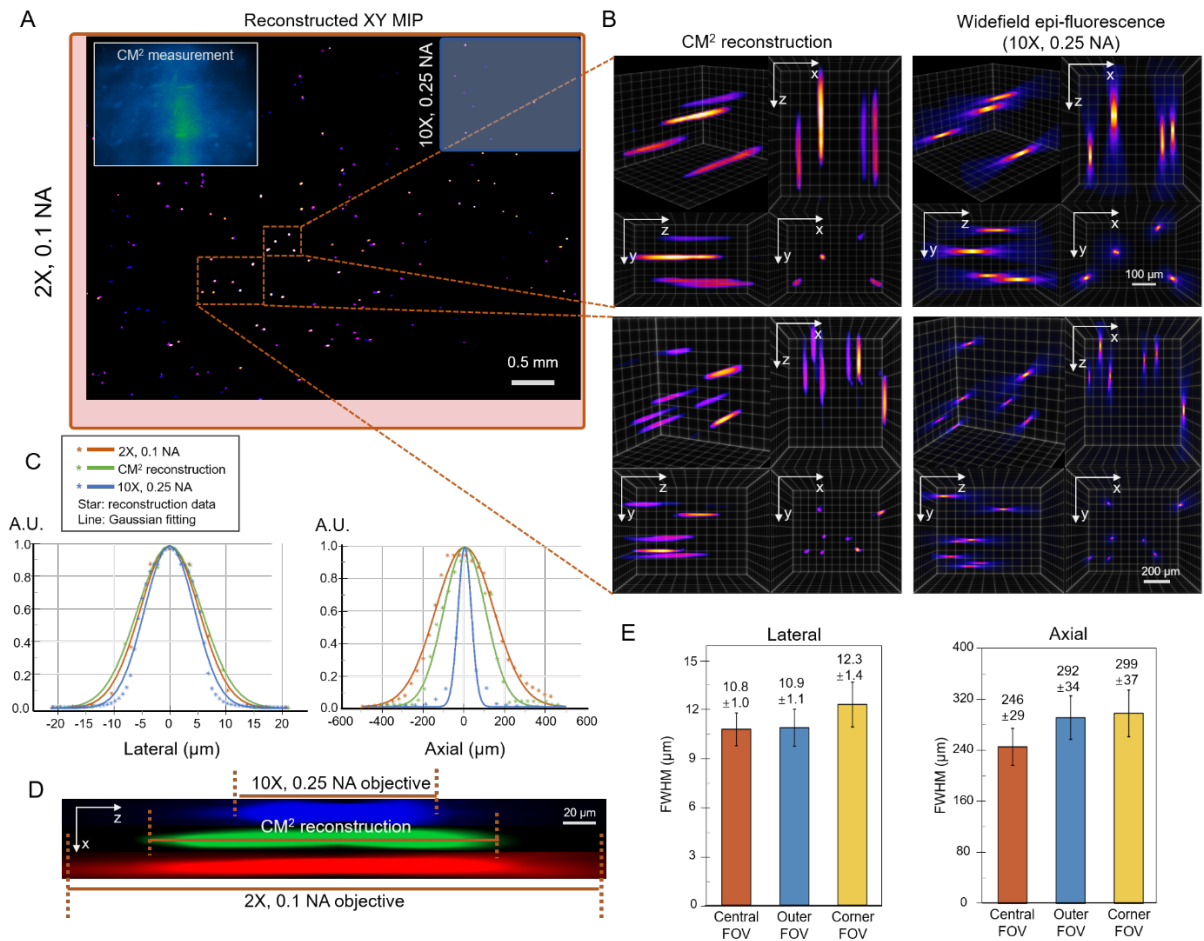

**Fig. 3. Single-shot 3D imaging of 10-μm fluorescent particles in a clear volume.** (**A**) XY MIP of the reconstructed volume spanning 5.7 mm × 6.0 mm × 1.0 mm. Top left insert: the raw CM² measurement. The FOV of the CM² is comparable to a 2× objective lens (red bounding box) and is ~25× wider than the 10× objective lens (blue bounding box). (**B**) Zoom-in of the CM² 3D reconstruction benchmarked by the axial stack taken by a 10×, 0.25 NA objective lens. (**C**) The lateral and axial cross-sections of the recovered 10-μm particle. By comparing with the measurements from the standard widefield fluorescence microscopy, the CM² faithfully recovers the lateral profile of the particle and achieves single-shot depth sectioning. (**D**) The XZ cross-sectional view of a reconstructed fluorescent particle, as compared to the axial-stack acquired from the 2× and 10× objective lenses. (**E**) To characterize the spatial variations of the reconstruction, the statistics of the lateral and axial FWHMs of the reconstructed particles are plotted for the central, outer and corner FOV (as defined in Fig. 2C). The lateral width changes only slightly (~0.9%) in the outer FOV but increases in the corner FOV (~13.9%). The axial elongation degrades from ~246 μm in the central FOV to ~292 μm and ~299 μm in the outer and corner FOV regions, respectively.

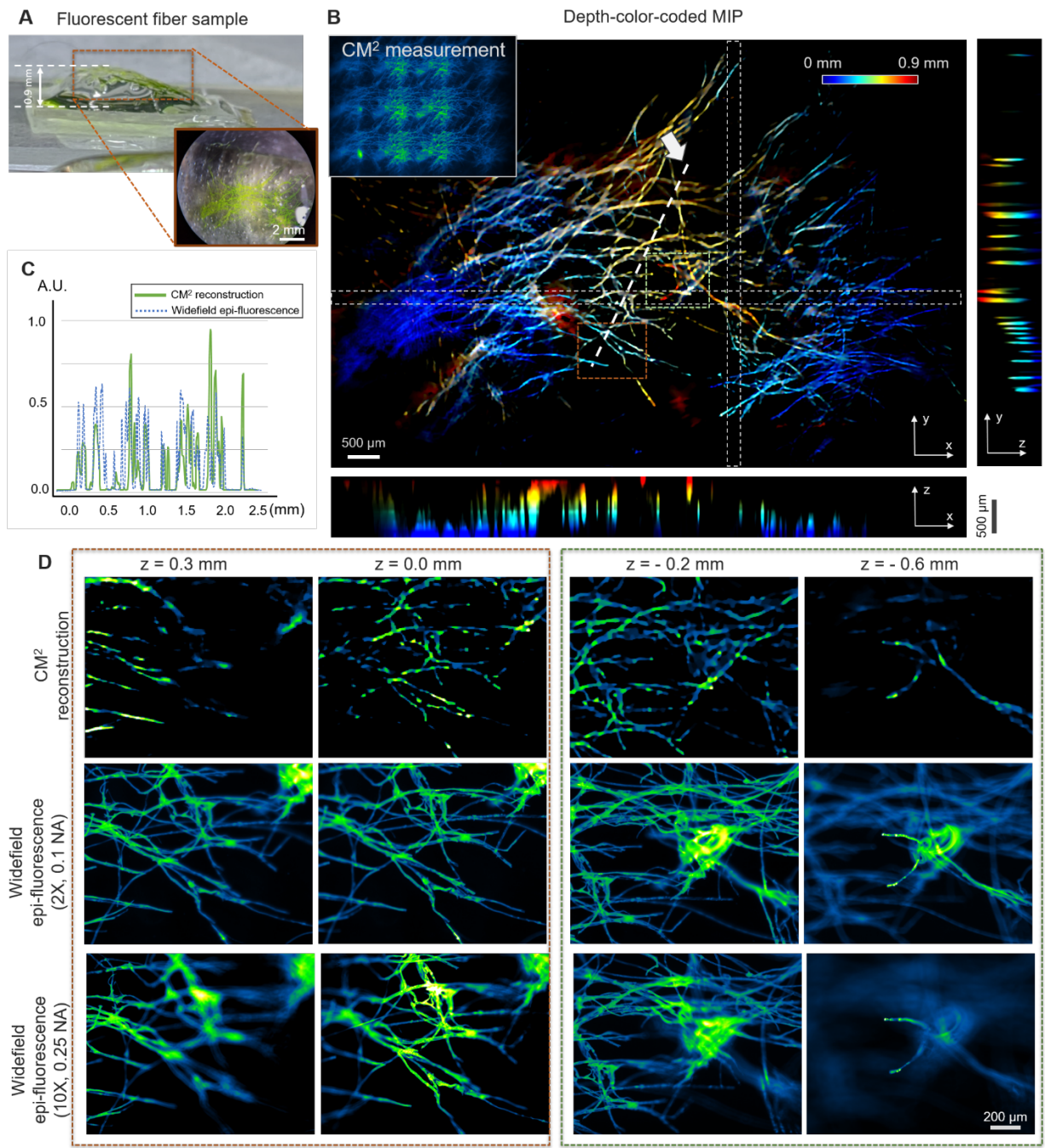

**Fig. 4. Imaging of fluorescent fibers on a curved surface.** **(A)** The sample contains fluorescent fibers spread on a 3D printed curved surface that mimics the mouse cortex. Photo Credit: Yujia Xue, Boston University. **(B)** The depth-color-coded MIP of the reconstruction spanning a volume of ~7.8 mm × 4.9 mm × 0.9 mm. The orthogonal projections reveal the curvature of the sample. **(C)** The $CM^2$ resolves the fiber structures as verified by the cutline from the reconstruction compared to the measurement with a 2× 0.1 NA objective lens. **(D)** The depth sectioning of the $CM^2$ is benchmarked by the widefield measurements from 2× 0.1 NA and 10× 0.25 NA objective lenses. The $CM^2$ accurately recovers in-focus fiber structures and suppresses out-of-focus fluorescence.

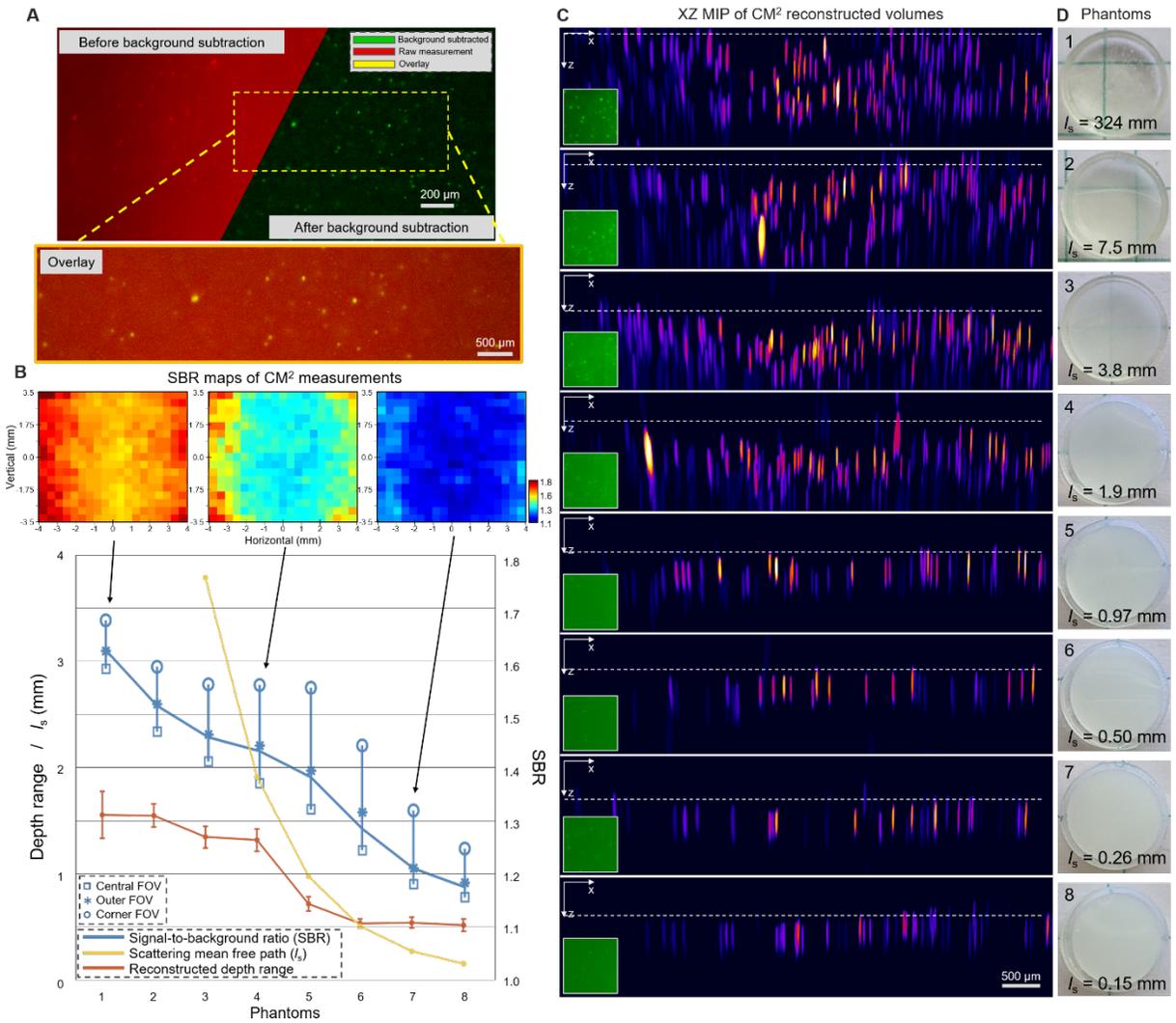

**Fig. 5. Imaging of scattering phantoms.** (**A**) A background subtraction procedure is devised to remove the slow varying background before performing the 3D deconvolution. (**B**) Quantitative evaluation of the $CM^2$ performance under bulk scattering and strong background fluorescence. The contrast of the raw measurement is quantified by the local SBR. The mean SBR across the whole FOV is plotted in the blue curve along with the mean SBR in different FOV regions (as defined in Fig. 2C). The SBR increases as the number of views in the measurement reduces. The FOV-dependent SBR is visualized by the local SBR maps at three scattering densities. The reconstructed depth range is measured to quantify the $CM^2$'s axial imaging capability. When the scattering is weak, the reconstructed depth range is primarily limited by the low SBR due to background fluorescence. As the scattering increases and $l_s$ approaches the axial elongation of a single particle, the reconstructed range reduces to the superficial layer and is bounded by the finite axial resolution. The error bars represent measurements from multiple sub-FOVs. $l_s$ for Phantom 1 and 2 are 323.7 mm, 7.5 mm, respectively, and are omitted in the plot for better visualization. (**C**) The XZ MIPs of the reconstruction across eight phantoms with different scattering densities. The dashed line in each sub-figure indicates the top surface of each phantom. (**D**) The images of the scattering phantoms used in the experiments. Photo Credit: Yujia Xue, Boston University.

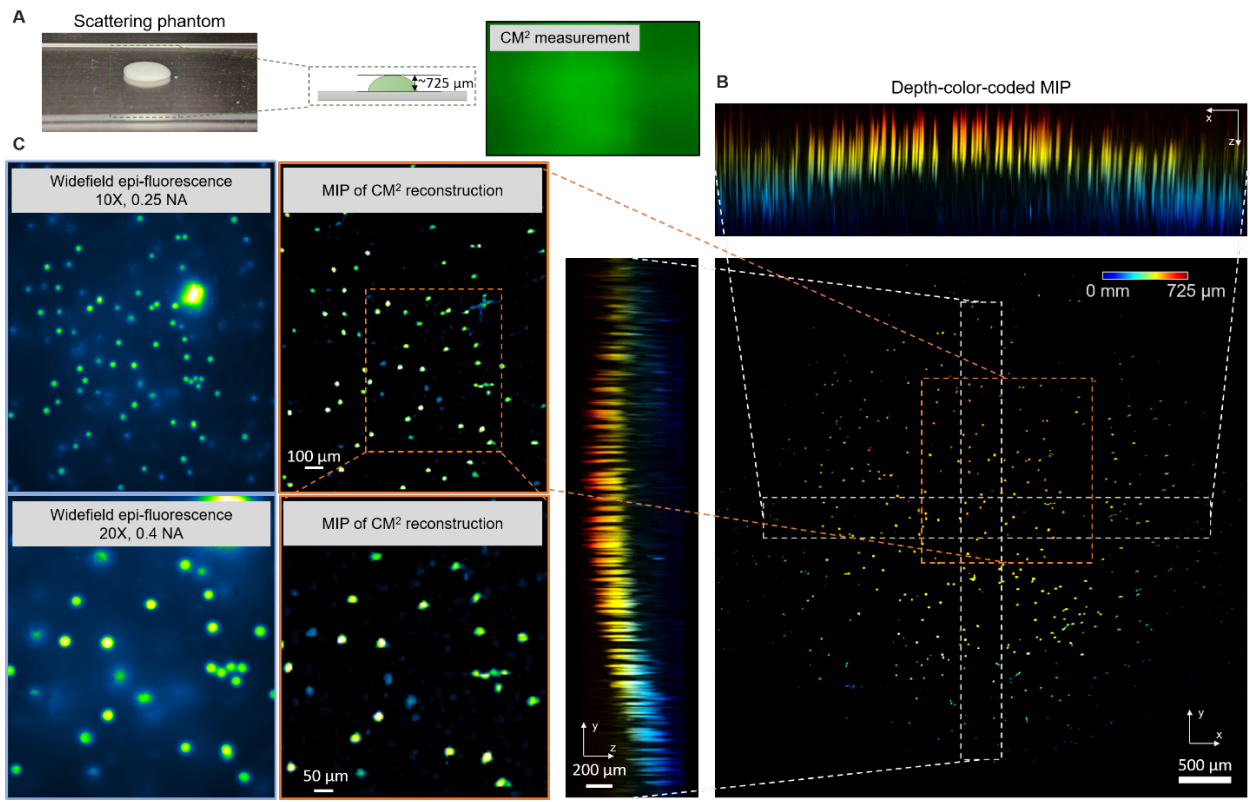

**Fig. 6. Imaging of a scattering sample with a curved surface.** (**A**) The illustration of the scattering sample ($l_s \sim 264$ μm) with a curved surface and the $CM^2$ raw measurement. Photo Credit: Yujia Xue, Boston University. (**B**) The depth-coded MIPs of the $CM^2$ reconstruction recovers particles in the superficial layer of the curved surface. (**C**) The comparison between the $CM^2$ reconstruction and the widefield fluorescence measurements (10×, 0.25 NA and 20×, 0.4 NA) verifies that the $CM^2$ correctly reconstructs the emitters in the superficial layer.